\documentclass[conference]{IEEEtran}

\RequirePackage[normalem]{ulem} 
\RequirePackage{color}\definecolor{RED}{rgb}{1,0,0}\definecolor{BLUE}{rgb}{0,0,1} 

\RequirePackage{color}

\usepackage{tikz}
\usepackage[sumlimits,intlimits]{amsmath}
\usepackage{amsmath,amsfonts,amssymb}%
\usepackage{bm}%
\usepackage{graphicx,graphics}%
\usepackage{cases}%
\usepackage[noadjust]{cite}%
\usepackage{color}%
\usepackage{cite,url}
\usepackage{verbatim}
\usepackage{enumerate}
\usepackage{algorithm}
\usepackage{algorithmic}
\usepackage{balance}
\usepackage{multirow}
\usepackage{stfloats}
\usepackage{subfigure}
\usepackage{xspace}
\usepackage{smartref}

\begin{document}

\title{\vspace*{-14pt}Overloaded Satellite Receiver Using SIC with Hybrid Beamforming and ML Detection
}
\author{\vspace*{0pt}
\authorblockN{Zohair Abu-Shaban$^\dag$, Hani Mehrpouyan$^\ddag$, Joel Grotz$^\S$ and Bj\"{o}rn Ottersten$^\dag$.\vspace*{-10pt}}
\IEEEauthorblockA{\\$^\dag$ SnT -- University of Luxembourg, Luxembourg.\\
$^\ddag$ Department of ECE\&CS, California State University, Bakersfield, CA, USA.\\
$^\S$ SES S. A., Betzdorf, Luxembourg.\\
emails: zohair.abushaban@uni.lu, hani.mehr@ieee.org, jgrotz@ieee.org, bjorn.ottersten@uni.lu.
\vspace{-10pt}
}}
\maketitle
\thispagestyle{empty}

\begin{abstract}


In this paper, a new receiver structure that is intended to detect the signals from multiple adjacent satellites in the presence of other interfering satellites is proposed. We tackle the worst case interference conditions, i.e., it is assumed that uncoded signals that fully overlap in frequency arrive at a multiple-element \emph{small-size} parabolic antenna in a spatially correlated noise environment. The proposed successive interference cancellation (SIC) receiver, denoted by \emph{SIC Hy/ML}, employs hybrid beamforming and disjoint maximum likelihood (ML) detection. Depending on the individual signals spatial position, the proposed SIC Hy/ML scheme takes advantage of two types of beamformers: a maximum ratio combining (MRC) beamformer and a compromised array response (CAR) beamformer. The performance of the proposed receiver is compared to an SIC receiver that uses only MRC beamforming scheme with ML detection for all signals, a joint ML detector, and a minimum mean square error detector. It is found that SIC Hy/ML outperforms the other schemes by a large margin.
\vspace{+3pt}

\end{abstract}

\section{Introduction}
The growth in the broadcasting satellites field warrants more satellites to be launched and stationed in space, usually in the geostationary orbit (GEO). For this reason, the frequency bands for broadcasting satellites, mainly Ku band, are densely occupied causing the satellite receivers performance to be significantly limited by \emph{adjacent satellite interference (ASI)} \cite{grotz2010}. Moreover, although the use of smaller receiver dishes by end users are more cost-effective from a commercial point of view, they possess wider beam patterns allowing more ASI at the receiver. These two reasons make the cancellation of ASI an urging need, with special attention paid to overloaded scenarios\footnote{More interfering signals than receiver antenna elements \cite{grotz2010}.} that conventional receivers can hardly cope with. The ability to receive from different adjacent satellites at the same time using a small-size antenna has the potential of increasing the system throughput.

{\let\thefootnote\relax\footnotetext{{
This work was supported by the National Research Fund (FNR), Luxembourg.
\vspace{-10pt}
}} }

Different techniques for multiuser detection in overloaded systems are found in the literature, e.g., \cite{hicks2001,kapur2003,Colman2008,krause2011}. Interference cancellation for satellite systems with signals partially overlapping in frequency is considered in \cite{beidas2002}, where signals are assumed to overlap partially in frequency allowing higher spectral efficiency. Subsequently, based on this assumption, interference cancellation for coded and uncoded data is studied \cite{beidas2002}. The work in \cite {beidas2002} is extended to the first and second generations of digital video broadcasting standards, DVB-S and DVB-S2, in \cite{Schwarz2007}. However, the contributions in \cite{beidas2002} and \cite{Schwarz2007} do not apply any spatial processing at the receiver, ignore useful spatial information that could be exploited at the receiver and, as such, are only applicable to antennas equipped with a single receiver element.

The first application of interference cancellation to broadcast reception has been considered in \cite{grotz2010}. This application uses \emph{multiple low noise blocks (MLNBs)}, i.e., multiple feeds, in a two-stage overloaded receiver. The first stage is a beamformer that spatially pre-conditions the received signals before time-domain detection. This is necessary because linear conventional receivers perform weakly in an overloaded setup \cite{honig2008}. However, the algorithm in \cite{grotz2010} cannot support the scenario where the neighboring satellites operate in the same frequency channel. This is especially important since the current satellite systems need to employ the same bandwidth to meet the throughput requirement for broadcast applications.



\begin{figure}[!t]
\begin{center}
  \includegraphics[scale=.8]{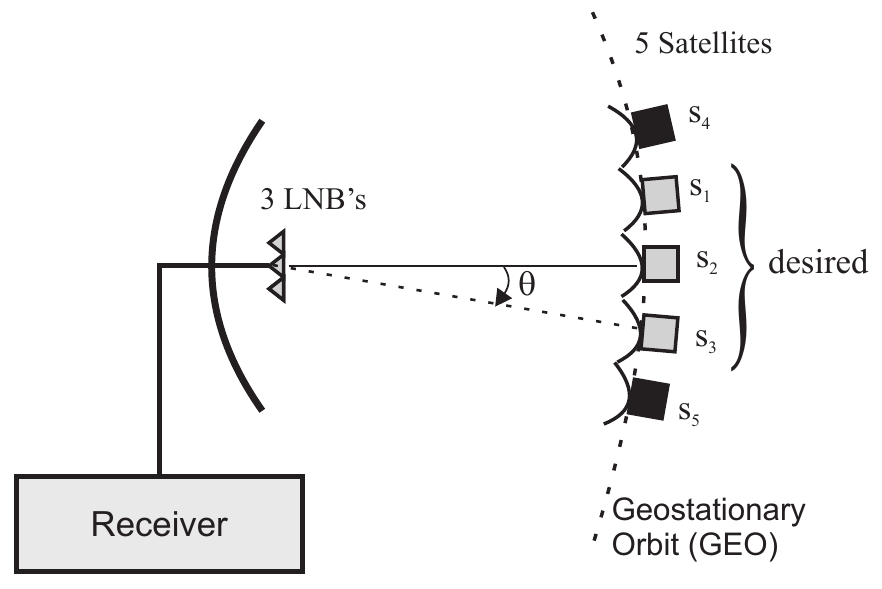}\\
  \caption{The scenario setup for $N_s=5$ satellites and $M=3$ LNBs. The dish is mainly directed to $s_2$.}\label{fig:setup}
  \end{center}
  \vspace{-.7cm}
\end{figure}

In this paper, we consider the problem of receiver design for detecting broadcast signals from adjacent satellites stationed in the GEO, when they are fully overlapping in the frequency. We assume uncoded transmission, where signals are assumed to be transmitted synchronously. The fixed receiver is assumed to be equipped with a small-size, ($<40$ cm), parabolic antenna (dish) equipped with multiple LNBs as illustrated in Fig. \ref{fig:setup}. The dish is assumed to be mainly directed towards one satellite, which we refer to as the \emph{main desired signal}. Including the main desired signal, the total number of signals we aim to detect is equal to the number of the LNBs used. The proposed receiver implements successive interference cancellation (SIC) using disjoint maximum likelihood (ML) detection after beamforming at the output of the respective LNB.\footnote{ML denotes disjoint maximum likelihood unless stated otherwise.} Two types of beamformers are used depending on the SIC iteration. This allows for a better utilization of the spatial properties of the received signal at each iteration. In the first iteration, the maximum ratio combining (MRC) beamformer, applied in satellite communications in \cite{grotz2010}, is used. MRC uses information about all the satellites to detect the main desired signal by maximizing the signal-to-interference-and-noise-ratio (SINR). The proposed compromised array response (CAR) beamformer is then used in the following iterations. The performance of the receiver is measured in terms of the bit error rate (BER) of the individual signals and the average BER for all detected signal. The contributions of this paper can be summarized as follows:

\begin{itemize}

  \item The CAR beamformer, which enhances the array response (AR) beamformer to better suit interference cancellation for satellite systems is proposed.

  \item By applying the proposed CAR beamformer, a hybrid receiver design denoted by \emph{SIC Hy/ML} is proposed to detect the signals from multiple satellites in the presence of interference from many neighboring satellites in an overloaded scenario. In contrast to \cite{grotz2010}, \cite{beidas2002}, and \cite{Schwarz2007}, which are based on the assumption that the signals from interfering satellites are only partially overlapping, here, we focus on the worst case scenario where all the interfering signals completely overlap in frequency.


   \item Extensive simulations are carried out to compare the performance of the proposed SIC Hy/ML approach against traditional MRC or AR based interference cancellers. These simulations demonstrate the superior BER performance of the proposed SIC Hy/ML receiver in mitigating interference compared to the MRC or AR approaches.
\end{itemize}

The rest of this paper is organized as follows: Section \ref{sec:section-sys-model} describes the system model, the scenario under consideration, and the assumptions in this work. Section \ref{sec:section-bf_ic} outlines the proposed beamformer and the detection technique, SIC Hy/ML. In Section \ref{sec:sim}, the simulation environment and results are presented, while Section \ref{sec:conc} concludes the paper.

\section{System Model and Assumptions}\label{sec:section-sys-model}
It is assumed that $N_s$ adjacent satellites stationed in GEO are broadcasting to a receiver equipped with a small-size parabolic dish with $M$ LNBs (see Fig. \ref{fig:setup}). The satellites are assumed to be within the field of view of the multiple-element reflector antenna. The considered system is assumed to be overloaded, i.e., $N_s >M$. The $N_s$ transmitted signals are assumed to be fully overlapping in frequency. Moreover, it is assumed that they are interfering synchronously at symbol time. This constitutes the worst case scenario in adjacent satellite interference situation. We assume that the satellites belong to one operator or to collaborating operators so that the system parameters, such as modulation, channel codes, and signal power are known. Finally, the signals are assumed to be compliant with the DVB-S2 standard \cite{dvbb-s2} and are independently transmitted, i.e., they are uncorrelated.

Under the assumption of perfect synchronization\footnote{Taking into account the effect of synchronization is subject of future research.}, the received signal vector at the output of the receiver analog-to-digital converter is modelled as:
\begin{align}\label{eq:sys-eq}
\mathbf{r}[k] = \mathbf{As}[k]+\mathbf{n}[k],
\end{align}
where $\textbf{r}[k]\triangleq\left[r_1 [k], r_2 [k], ..., r_M [k]\right]^T$ is the received symbols vector at a time instant $k$, $\textbf{A}\triangleq[a_{i,j}]$ is an $M\times{N_s}$ matrix representing the antenna array response with $a_{i,j}$ denoting the complex gain of the $i^{th}$ LNB in the direction of the $j^{th}$ satellite. $\textbf{s}[k]\triangleq\left[s_1 [k], s_2 [k], ..., s_{N_s}[k]\right]^T$ is the transmitted symbols vector and $\textbf{n}[k]\triangleq\left[n_1[k], n_2 [k], ..., n_M [k]\right]^T$ is the noise vector. A line-of-sight link is assumed. Hence, the channel mainly depends on the antenna geometrical specifications, i.e., diameter, focal length, etc. and LNBs electrical specifications, i.e., oscillator stability, low noise amplifier gain, etc. Since these parameters do not vary quickly, they are assumed to be fixed over the transmission period. Accordingly, ignoring pointing errors,the antenna radiation patterns, which $\textbf{A}$ is dependent on, are considered known and fixed. The noise is assumed spatially correlated since the radiation patterns of the MLNB's overlap causing one LNB noise pattern to affect the neighboring LNBs. Thus, the noise vector is modeled using an additive Gaussian noise with covariance matrix $\textbf{R}_n=\sigma^2\textbf{K}$, where $\sigma^2$ is the noise power and $\textbf{K}$ is the spatial correlation matrix. Each satellite is positioned at an angle $\theta_j$ degrees measured on the plane containing the MLNBs axis and GEO (see Fig. \ref{fig:setup}).


\section{The Proposed Receiver Design}\label{sec:section-bf_ic}
Three factors make the considered scenario challenging:
\begin{itemize}
  \item Firstly, fully overlapping signals occupy the same spectral area. There will be no areas where partial information can be exploited for interference cancellation.
  \item Secondly, small antenna requirement means less antenna pattern directivity, consequently, higher interference power.
  \item Finally, overloaded system implies that $\textbf{A}$ is a wide matrix, i.e., we have more variables to detect ($N_s$ signals) than equations ($M$ signals) in the linear system in (\ref{eq:sys-eq}).
\end{itemize}
For these reasons, a non-linear receiver is required. The overloaded SIC receiver proposed herein consists of two processes (1) Hybrid beamforming scheme that exploits the spatial features of the setup to reduce the overloading effect and (2) ML detection that estimates the symbols disjointly in the time domain. The proposed receiver is discussed in Subsection \ref{subsubsec:SIC_ML}.
\subsection{Beamforming}
Different performance metrics can be used to select the beamforming weighting parameters:
\subsubsection{MRC Beamformer}
Let $\mathbf{A}\triangleq[\mathbf{a}_1,\mathbf{a}_2,...\mathbf{a}_{N_s}]$, then dropping the time dependency $k$, we can rewrite (\ref{eq:sys-eq}) as
\vspace{-1pt}
\begin{align}\label{eq:receive}
 \mathbf{r}=\sum_{m=1}^{N_s}\mathbf{a}_{m}s_{m}+\mathbf{n}.
\end{align} \vspace{-1pt}
Given that the signals are uncorrelated, then the auto-covariance matrix for $\textbf{r}$ is given by
\begin{align}\label{eq:autocovariance_matrix}
 \mathbf{R}=\sum_{m=1}^{N_s}\mathbf{a}_{m}s_{m}s_{m}^{H}\mathbf{a}_{m}^H+\mathbf{R}_n=\sum_{m=1}^{N_s}\mathbf{R}_m+\mathbf{R}_n,
\end{align}
where $(.)^H$ is the complex conjugate transpose. The MRC beamformer for the $m^{th}$ signal is given by
\begin{align}\label{eq:MRC_BF}
 \mathbf{w}_m=arg \max_{w}\frac{\mathbf{w}^H\mathbf{R}_m\mathbf{w}}{\mathbf{w}^H(\mathbf{R}-\mathbf{R}_m)\mathbf{w}}.
\end{align}
This is a generalized Rayleigh quotient whose optimization is a generalized eigenvalue problem. Thus, the solution $\textbf{w}_m$ is the eigenvector corresponding to the greatest eigenvalue of $(\mathbf{R}-\mathbf{R}_m)^{-1}\mathbf{R}_m$, given the inverse exists \cite{eigen}.

\subsubsection{AR beamformer}
In the vector space containing $\{\mathbf{a}_1,\mathbf{a}_2,...,\mathbf{a}_{N_s}\}$, the AR beamformer is a scaled vector in the direction of the array response vector of the signal to be detected. The AR beamformer to detect the $m^{th}$ signal is given by\vspace{-6pt}
\begin{align}\label{eq:AR_BF}
 \mathbf{w}_m=\frac{\mathbf{a}_m}{\|\mathbf{a}_m\|^{2}}.
\end{align}
In contrast to the MRC beamformer, the AR beamformer is easy to construct as it depends only on the direction and does not require the calculation of the auto-covariance matrices.

\subsection{Detection Techniques}
In the proposed receiver, SIC is applied with hybrid beamforming and ML detection (SIC Hy/ML). By applying a hybrid scheme, we are able to apply a more appropriate beamformer after each SIC iteration, where the number of interferes is reduced. Comparison against \cite{honig2008} and \cite{Verdu:1998:MD:521411} shows that the proposed scheme outperforms the unified beamforming approaches for all signals.

\subsubsection{Joint Maximum Likelihood Detector (JML)}
The choice of the JML detector is motivated by its optimality for equiprobable transmitted symbols \cite{honig2008}. Writing $\textbf{s}$ and $\textbf{A}$ in terms of their desired and interference parts, we have, $\mathbf{s}=[\mathbf{s}_d^T,\mathbf{s}_i^T]^T$ and $\mathbf{A}\triangleq[\mathbf{a}_1,\mathbf{a}_2,...\mathbf{a}_{N_s}]=[\mathbf{A}_d,\mathbf{A}_i]$. Note that $\textbf{s}_d$ and $\textbf{s}_i$ contain the desired and interfering signals, respectively. Within $\textbf{s}_d$ and $\textbf{s}_i$, the signals are ordered according to their position with respect $\theta=0$ (see Fig. \ref{fig:setup}). $\textbf{A}_d$ and $\textbf{A}_i$ are $M\times{M}$ and $M\times({N_s}-M)$ matrices, respectively.

Let the output of the beamformer be given by
\begin{align}
\mathbf{p}=\mathbf{W}^{H} \mathbf{r}=\mathbf{W}^{H}\mathbf{A}_d\mathbf{s}_d+\mathbf{W}^{H}\mathbf{A}_i\mathbf{s}_i+\mathbf{W}^{H}\mathbf{n},
\end{align}
where, $\mathbf{W}=\left[\mathbf{w}_1,\mathbf{w}_2,...\mathbf{w}_M\right]$.
Then, the JML detector is determined as
\begin{align}
\hat{\mathbf{s}}_d=arg \min_{\mathbf{s}_d} {\|\mathbf{p}-\mathbf{W}^{H}\mathbf{A}_d\mathbf{s}_d\|}^2.
\end{align}
\subsubsection{Minimum Mean Square Error (MMSE) Detector}
The desired transmitted vector is linearly estimated using an $M\times{M}$ matrix $\textbf{C}$ such that\vspace{-6pt}
\begin{align}
\hat{\mathbf{s}}_d=\textbf{C}\textbf{p},
\end{align}
where $\mathbf{C}= arg \min_{\mathbf{C}_{o}} {\|\mathbf{s}_{d}-\mathbf{C}_{o}\mathbf{p}\|}^2$. Following a similar procedure to the one found in \cite{Verdu:1998:MD:521411}, it is easy to show that\vspace{-6pt}
\begin{align}
 \mathbf{C}= \mathbf{R}_d \mathbf{A}_d^H\mathbf{R}^{-1}(\mathbf{W}^H)^{-1},
\end{align}
where $\mathbf{R}_d=\sum_{m=1}^{M}\mathbf{R}_m$. Writing $\hat{\mathbf{s}}_d$ as a function of $\mathbf{r}$, we have
\begin{align}\label{eq:W-Gone}
\hat{\mathbf{s}}_d=\left(\mathbf{R}_d \mathbf{A}_d^H\mathbf{R}^{-1}(\mathbf{W}^H)^{-1}\right)\left(\mathbf{W}^{H}\mathbf{r}\right)=\mathbf{R}_d \mathbf{A}_d^H\mathbf{R}^{-1}\mathbf{r}.
\end{align}
Note that according to \eqref{eq:W-Gone}, the beamforming does not change the performance of the MMSE detector.
\subsubsection{Successive interference cancellation with hybrid beamforming and ML detection (SIC Hy/ML)}\label{subsubsec:SIC_ML}
This hybrid iterative detection technique uses an MRC beamformer in the first iteration in order to isolate the main desired signal. Subsequently, the proposed CAR beamformer is used in the following iterations to isolate the remaining $M-1$ desired signals. Since the position of each satellite in GEO is fixed and known, these beamformers can benefit from this spatial knowledge to preprocess the received signals by shaping the antenna pattern of the LNBs towards their respective desired signal. Without loss of generality, assume $M$ is odd so that the main desired signal is $s_{\frac{M+1}{2}}$ (otherwise, take the satellite closest to $\theta=0^{0}$ as the main desired signal). Constructing the MRC beamformer using (\ref{eq:MRC_BF}) for the main desired signal, then the output of the beamformer can be determined as
\begin{align}\label{eq:BF_output}
p_{\frac{M+1}{2}}=\mathbf{w}_{\frac{M+1}{2}}^{H}\mathbf{r}=\sum_{m=1}^{N_s}\mathbf{w}_{\frac{M+1}{2}}^H\mathbf{a}_{m}s_m+\mathbf{w}_{\frac{M+1}{2}}^H\mathbf{n}.
\end{align}
Subsequently, the ML detector can be determined as
\begin{align}\label{eq:s2_estimate}
\hat{{s}}_{\frac{M+1}{2}}=arg \min_{{s}_{\frac{M+1}{2}}} {\|p_{\frac{M+1}{2}}-\mathbf{w}_{\frac{M+1}{2}}^{H}\mathbf{a}_{\frac{M+1}{2}}{s}_{\frac{M+1}{2}}\|}^2.
\end{align}
The complexity of this ML estimator is not high, given that the common broadcasting standards, e.g., DVB-S2, use low order modulation such as quadrature phase shift keying (QPSK), and $8$-phase shift keying \cite{dvbb-s2}. Consequently, the search in \eqref{eq:s2_estimate} is only for a scalar value within a small constellation.

When the main desired signal is detected, its effect can be removed from the received signal $\mathbf{r}$. For each of the desired signals, the output of the corresponding beamformer has a similar form to that of (\ref{eq:BF_output}). Note that the coefficient of $s_{m}$ is the inner product $\mathbf{w}^{H}\mathbf{a}_{m}$, where $\mathbf{w}$ is the beamformer used. For this reason, the AR beamformer constructed using $\mathbf{a}_{m}$ will maximize the power of the desired signal and reduce the power of the interference. However, due to the small separation between GEO satellites (usually $2.5^\circ$ - $3.0^\circ$)\footnote{$(\cdot)^\circ$ is used to denote degrees.}, isolating them by AR-beamforming may also amplify the interfering signals due to the wide beams of the AR beamformer. Thus, after removing the interfering signals, the proposed beamforming approach, denoted by compromised AR, constructs the beamformer in the direction of $\theta$ to maximize the difference between the coefficients of the desired signal and the signal from the closest satellite in (\ref{eq:receive}). The car beamformer for the $n^{th}$ signal can be written as
\begin{align}\label{eq:CAR_BF}
 \mathbf{w}_{c,n}(\hat{\theta}_{c,n})=\frac{\mathbf{a}(\hat{\theta}_{c,n})}{\|\mathbf{a}(\hat{\theta}_{c,n})\|^{2}},
\end{align}
where
\begin{align}\label{eq:theta_estimate}
\hat{\theta}_{c,n}=arg \max_{\theta} \left(\left|\mathbf{w}^H_{c,n}(\theta)\mathbf{a}_n\right|-\left|\mathbf{w}^H_{c,n}(\theta)\mathbf{a}_i\right|\right),
\end{align}
and $\mathbf{a}_i$ is the array response vector of the closest satellite. The proposed SIC Hy/ML continues detecting the subsequent signals using the CAR beamformer, $\textbf{w}_{c,n}(\theta), 1\leq{n}\leq{M}, {n}\neq{\frac{M+1}{2}}$, such that
\begin{align}\label{eq:SIC}
p_{n}=\sum^{M}_{m=n,m\neq{\frac{M+1}{2}}}{\mathbf{w}_{c,n}^{H}\mathbf{a}_{m}s_m}+\mathbf{w}_{c,n}^{H}\mathbf{n},
\end{align}
\vspace{-10pt}
\begin{align}\label{eq:s_estimate_car}
\hat{{s}}_{n}=arg \min_{{s}_{n}} {\|p_{n}-\mathbf{w}_{c,n}^{H}\mathbf{a}_{n}{s}_{n}\|}^2.
\end{align}
Note that since the summation in (\ref{eq:SIC}) starts from $m=n$ and skips $m=\frac{M+1}{2}$, the interfering signals are implicitly subtracted from the received signal. Algorithm 1 bellow summarizes the proposed SIC Hy/ML approach.
\vspace{-6pt}
\begin{algorithm}
\begin{algorithmic}[1]
\STATE \footnotesize Input  $\mathbf{r},  \mathbf{A}, \mathbf{K}, M, N_s, E_s, \sigma^{2}$.
\vspace{2pt}
\STATE \footnotesize Calculate $\mathbf{R}$ using (\ref{eq:autocovariance_matrix}).
\vspace{2pt}
\STATE \footnotesize {Find $\mathbf{w}_{\text{MRC}}$ for ${s}_{\frac{M+1}{2}}$, the eigenvector corresponding to the greatest\\
\vspace{2pt}
\STATE eigenvalue of $(\mathbf{R}-\mathbf{R}_{\frac{M+1}{2}})^{-1}$}.
\vspace{2pt}
\STATE \footnotesize $p_{\frac{M+1}{2}}\leftarrow\mathbf{w}_{\text{MRC}}^H\mathbf{r}$.
\vspace{2pt}
\STATE \footnotesize Find $\hat{{s}}_{\frac{M+1}{2}}$ using (\ref{eq:s2_estimate}).
\vspace{2pt}
\STATE \footnotesize Demodulate ($\hat{{s}}_{\frac{M+1}{2}}$).
\vspace{2pt}
\STATE \footnotesize $\mathbf{\grave{r}}\leftarrow\mathbf{r}-\mathbf{a}_{\frac{M+1}{2}}\hat{{s}}_{\frac{M+1}{2}}$.
\vspace{2pt}
\FORALL {$n=1$ to $n=M$ and $n\neq{\frac{M+1}{2}}$}
\STATE \footnotesize Find $\hat{\theta}_{c,n}$ from (\ref{eq:theta_estimate}).
\vspace{2pt}
\STATE \footnotesize Construct $\mathbf{w}_{c,n}$ from (\ref{eq:CAR_BF}).
\vspace{2pt}
\STATE \footnotesize $p_{n}\leftarrow\mathbf{w}_{c,n}^H\mathbf{\grave{r}}$.
\vspace{2pt}
\STATE \footnotesize Find $\hat{{s}}_{n}$ using (\ref{eq:s_estimate_car}).
\vspace{2pt}
\STATE \footnotesize Demodulate ($\hat{{s}}_{n}$).
\vspace{2pt}
\STATE \footnotesize$\mathbf{\grave{r}}\leftarrow\mathbf{\grave{r}}-\mathbf{a}_{n}\hat{{s}}_{n}$.
\ENDFOR
\caption{SIC Hy/ML description}
\end{algorithmic}
\end{algorithm}

\vspace{0pt}
\section{Simulation Results and Discussion}\label{sec:sim}
\begin{figure}[!t]
\begin{center}
\vspace{-.53cm}
  \includegraphics[scale=0.78]{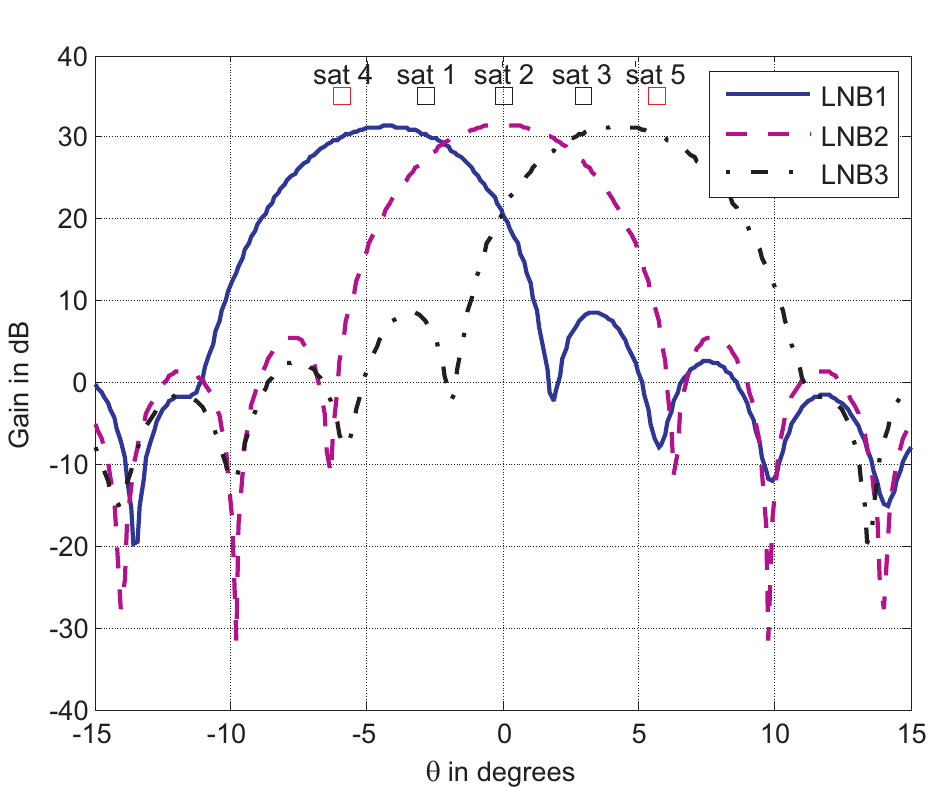}\\
      \vspace{-.3cm}
  \caption{Antenna patterns of LNBs $1$, $2$ and $3$ fixed on a 35-cm parabolic dish. Satellites depicted in black are desired while the ones in red are interferers.}\label{fig:patterns}
  \end{center}
  \vspace{-0.7cm}
\end{figure}

\vspace{0.2cm}
In this section, the proposed receiver's BER performance is investigated. Monte-Carlo simulation results for the setup depicted in Fig \ref{fig:setup}, which consists of 5 satellites and 3 LNBs ($N_s=5$, $M=3$), are discussed. In this simulation setup the receiver is assumed to be equipped with a $35$ cm diameter dish antenna. The antenna is directed towards the satellite transmitting the main desired signal $s_2$. The $5$ satellites are stationed at geostationary angles $-2.8^\circ, 0^\circ, 3^\circ, -5.9^\circ$ and $5.7^\circ$, respectively. According to the DVB-S2 standard, QPSK modulation is deployed. The noise power is calculated via $\sigma^2=\frac{E_s\|A\|^2_{F}}{\text{SNR}\times{M}}$, where $\|.\|_{F}$ is the Frobenius norm, $E_s=1$ is the symbols energy, and SNR denotes the signal-to-noise ratio \cite{Colman2008}. The software package \emph{GRASP} is used to obtain the radiation patterns of the three LNBs, see Fig. \ref{fig:patterns}. Due to the small diameter of the dish, the antenna directivity is low and the antenna patterns are wide, resulting in a high level of interference. The noise spatial correlation matrix is estimated as 
\begin{align} \label{eq:K}
\mathbf{K}=
\left(
  \begin{array}{ccc}
    1& 0.1& 0.05\\0.1& 1 &0.1\\ 0.05& 0.1& 1\\
  \end{array}
\right).
\end{align}
Exact values for the $i$th row and $j$th column entry of the matrix $\mathbf{K}$ in \eqref{eq:K} can be obtained by normalizing the intersection area under the antenna gain plots of LNB$_i$ and LNB$_j$ in Fig. \ref{fig:patterns}.
\begin{figure}[!b]
\begin{center}
\vspace{-.6cm}
  \includegraphics[scale=0.78]{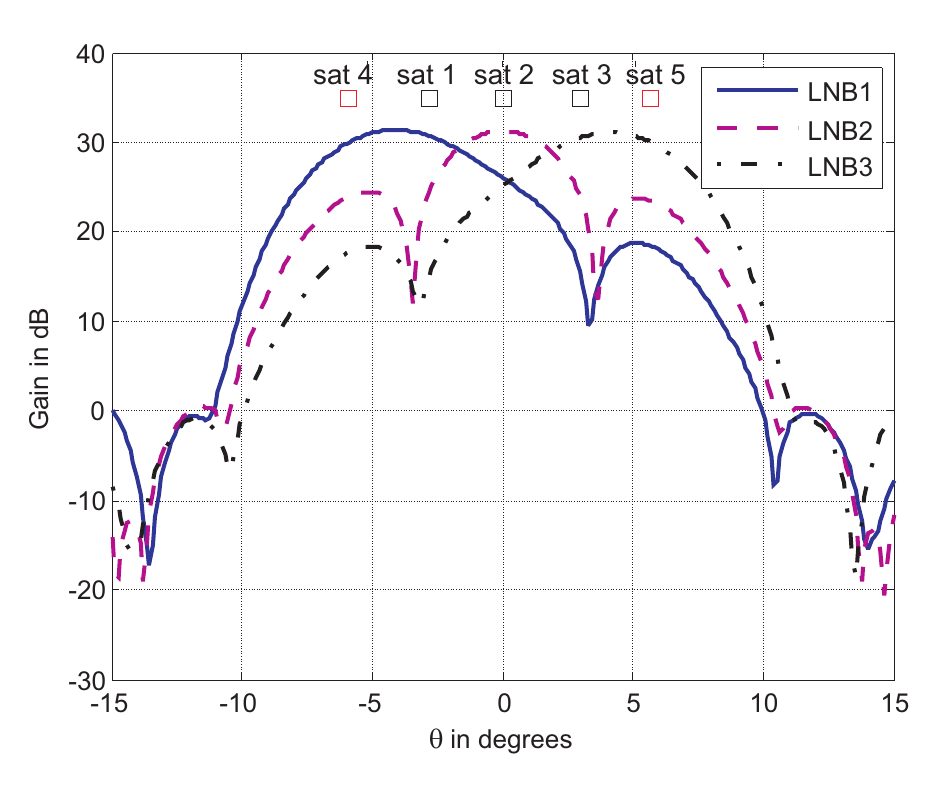}\\
    \vspace{-.4cm}
  \caption{Antenna patterns of LNBs $1$, $2$ and $3$ after using the MRC beamformer.}\label{fig:MRC_pattern}
  \end{center}
  \vspace{-0cm}
\end{figure}
\begin{figure}[!t]
\begin{center}
\vspace{-.5cm}
  \includegraphics[scale=0.78]{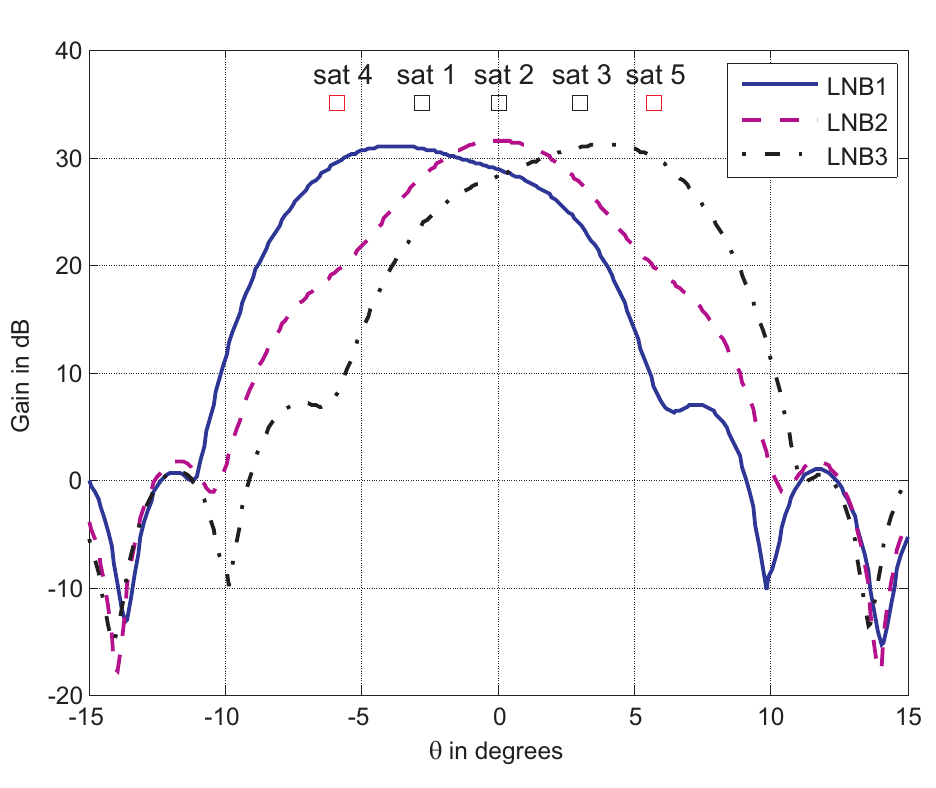}\\
    \vspace{-.4cm}
  \caption{Antenna Patterns of LNBs $1$, $2$ and $3$ after using AR beamformers.}\label{fig:AR_pattern}
  \end{center}
  \vspace{-0.6cm}
\end{figure}
\begin{figure}[!b]
\begin{center}
\vspace{-.5cm}
  \includegraphics[scale=0.78]{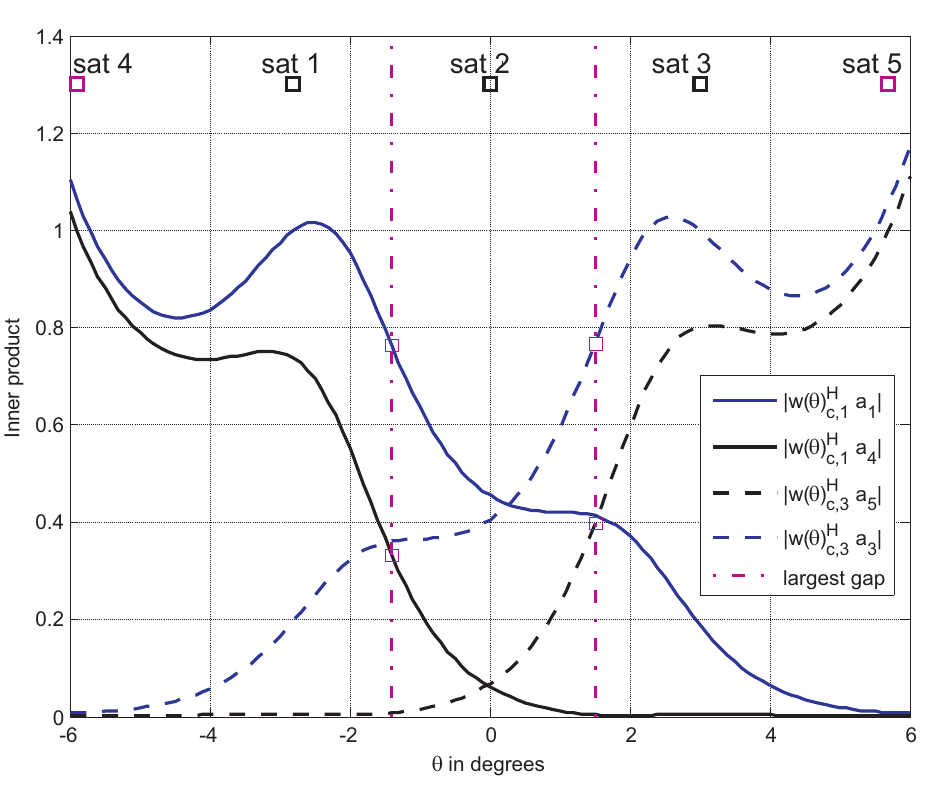}\\
    \vspace{-.3cm}
  \caption{The graphical solution of (\ref{eq:theta_estimate}), $\hat{\theta}_{c,1}=-1.4^\circ$ and $\hat{\theta}_{c,3}=1.5^\circ$).}\label{fig:bestBF_1}
  \end{center}
  \vspace{-0cm}
\end{figure}
\subsection{MRC and AR beamformers}\label{sec:mrc_sim_results}
Fig. \ref{fig:MRC_pattern} illustrates the MRC beamformers for the three LNBs. The MRC beam for the signal $s_2$ is sharp and constitutes a good channel that can reject $s_1$, $s_3$, $s_4$, and $s_5$. However, the beams for $s_1$ and $s_3$ allow adjacent satellites to interfere with their respective desired signal, namely, $s_4$ and $s_5$, respectively. The AR beams are shown in Fig. \ref{fig:AR_pattern}, where it can be seen that the AR beams are wider than their MRC counterparts. For this reason, MRC is used in the first iteration of the proposed SIC Hy/ML algorithm and for their steerability, CAR beamformers are used in the subsequent iterations when the SINR improves due to application of SIC.
\subsection{CAR beamformer}

\begin{figure}[!t]
\begin{center}
  \includegraphics[scale=0.78]{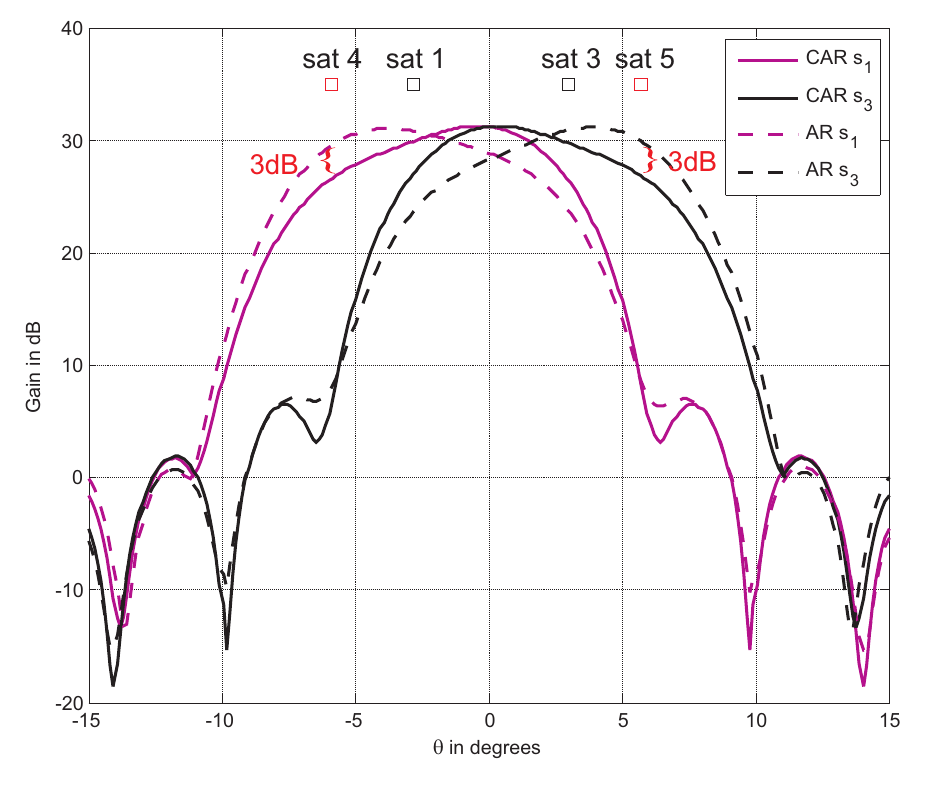}\\
    \vspace{-.4cm}
  \caption{A comparison between the AR and CAR beamformers to detect $s_1$ and $s_3$. $s_1$ is detected after canceling $s_2$, while $s_3$ is detected after canceling $s_1$ and $s_2$. $\theta_{AR,1}=-2.8^\circ$, $\hat{\theta}_{c,1}=-1.4^\circ$, $\theta_{AR,3}=3^\circ$ and $\hat{\theta}_{c,3}=1.5^\circ$.}\label{fig:AR_CAR_1}
  \end{center}
  \vspace{-0.6cm}
\end{figure}

After removing $s_2$ in the previous iteration via the MRC beamformer, Fig. \ref{fig:bestBF_1} is used to solve (\ref{eq:theta_estimate}) graphically for $\hat{\theta}_{c,1}$, given that $s_4$ is the closest satellite. To reduce the gain of $s_4$, $\hat{\theta}_{c,1}$ is expected to be between $s_1$ and $s_2$. Thus, the solution could be restricted to this range. It can be inferred that $\hat{\theta}_{c,1}=-1.4^\circ$, when the largest gap between $\left|\mathbf{w}^H_{c,1}(\theta)\mathbf{a}_1\right|$ and $\left|\mathbf{w}^H_{c,1}(\theta)\mathbf{a}_4\right|$ occurs. Using a similar approach $\hat{\theta}_{c,3}$ can be determined, given that $s_5$ in the closest satellite and that $s_1$ and $s_2$ have been already detected. In the scenario considered here, $\hat{\theta}_{c,3}$ is calculated to be $1.5^\circ$. The improvement offered by the CAR beamformer over AR beamformer is depicted in Fig. \ref{fig:AR_CAR_1}. This figure shows that by using CAR instead of AR, the interference due to $s_4$ and $s_5$ is reduced by $3$ dB.
\subsection{SIC Hy/ML}
\begin{figure}[!t]
\begin{center}
  \includegraphics[scale=0.78]{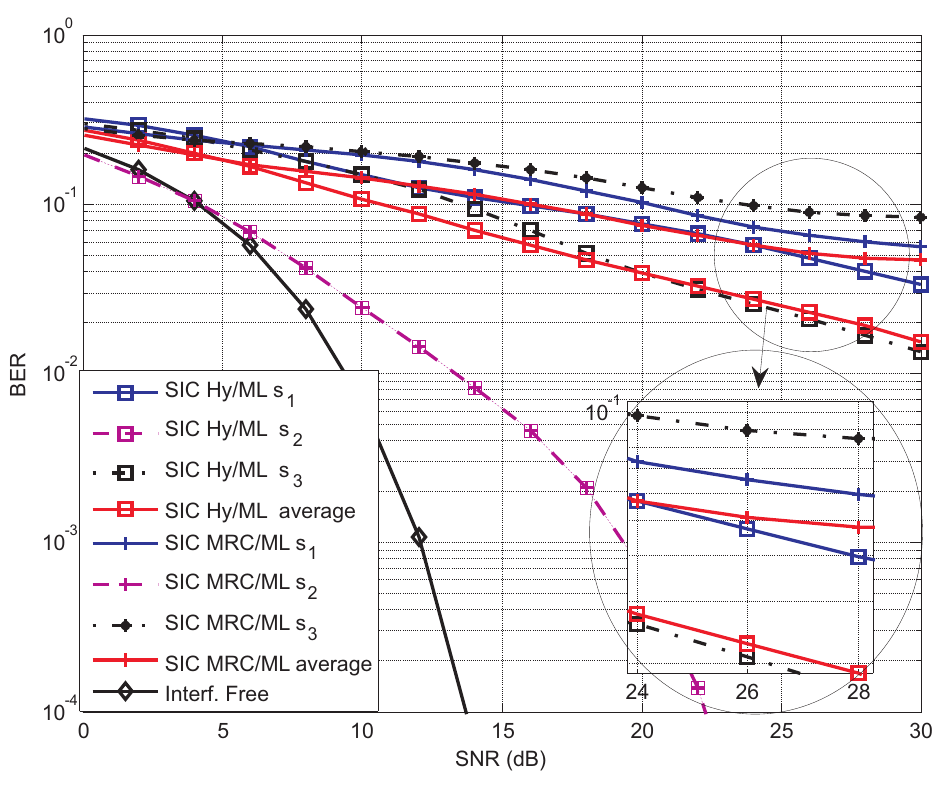}\\
    \vspace{-.4cm}
  \caption{BER for a SIC receiver with MRC beamformer and ML detector and a SIC receiver with hybrid beamforming and ML detector.}\label{fig:MRC_HY_together2}
  \end{center}
  \vspace{-0.6cm}
\end{figure}

Fig. \ref{fig:MRC_HY_together2} illustrates the BER curves for $s_1$, $s_2$, $s_3$ and the average (computed over these three signals) detected using SIC Hy/ML and SIC MRC/ML. Furthermore, this figure shows the lower bound for the BER performance, i.e., the interference-free scenario, in which all interferers are set to zero while keeping the setup unchanged. As indicated in \ref{sec:mrc_sim_results}, since $s_2$ has the best channel condition, it results in the best BER performance amongst the three received signals. Note that the proposed SIC Hy/ML and SIC MRC/ML performance for $s_2$ overlap, since in both approaches $s_2$ is detected by an MRC beamformer in the first iteration. Recall that $s_2$ and its strongest interferers, namely $s_1$ and $s_3$, are averaged in the average BER curve. It can be seen that the overall performance of SIC Hy/ML is better than SIC MRC/ML by a margin of $12$ dB. Furthermore, we note that $s_2$ can achieve BER performance adequate for quasi error free (QEF) reception for SNRs (E$_s$/N$_0$) below 15 dB. This is in line with a typical link budget considered here. This results in about $10$--$14$ dB clear sky fade margin for broadcast applications with this particular small antenna size. For example, our calculations show that when applying DVB-S2 with QPSK modulation, the considered system can achieve QEF reception with a $3/4$ forward error correction (FEC) channel code rate \cite{dvbb-s2}.
\begin{figure}[!t]
\begin{center}
  \includegraphics[scale=0.9]{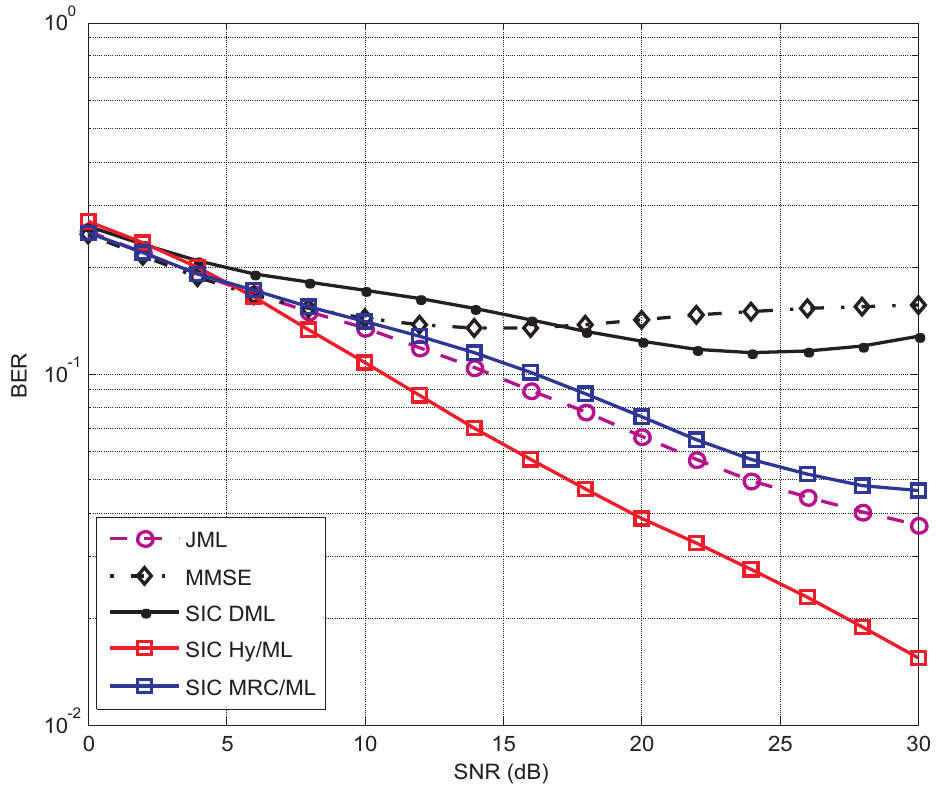}\\
    \vspace{-.4cm}
  \caption{A comparison of SIC Hy/ML performance to other detection methods.}\label{fig:all_detection}
  \end{center}
  \vspace{-0.6cm}
\end{figure}
In Fig. \ref{fig:all_detection}, the performance of the proposed SIC Hy/ML is compared to other detectors. The advantage of adapting the beamforming scheme to the individual signals by exploiting the spatial information available in the setup enables SIC Hy/ML to outperform JML. Although JML is the optimum joint detector, its performance barely improves with beamforming and hence, JML and MRC/JML are identical. While SIC MRC/ML and SIC Hy/ML detect $s_2$ with identical BER, there is a gap between their BER curves for the average scenario. This confirms the effect of using CAR to detect $s_1$ and $s_3$ on the overall BER. The SIC detector that exploits ML detection without beamforming is denoted by SIC DML in Fig. \ref{fig:all_detection} and can be considered as the upper bound for SIC as no beamforming is utilized before the ML detection. The MMSE is the worst among the applied detectors since it is a linear detector that performs badly in an overloaded condition \cite{Verdu:1998:MD:521411}.


\vspace{-6pt}
\section{Conclusions}\label{sec:conc}
In this paper, a receiver design for adjacent satellite interference cancellation is proposed. Three specific and practical system design issues made the problem under consideration challenging: (1) the signals being totally overlapping in frequency, (2) overloaded situation, and (3) the requirement of small dish antennas. Different beamformers (MRC and CAR) were adopted in the proposed receiver to facilitate interference cancelation and signal detection by $3$ dB. Moreover, simulation results show that the proposed SIC Hy/ML receiver outperforms the SIC MRC/ML, MMSE, and MRC/JML detectors by a large margin. Note that due to consideration of the complete frequency overlap, modest BER performances are observed. Nevertheless, for a link budget of $12$ dB, one can achieve the QEF requirement by a applying a forward error correcting code of rate $1/3$ with QPSK DVB-S2.



\begin{thebibliography}{10}
\providecommand{\url}[1]{#1}
\csname url@samestyle\endcsname
\providecommand{\newblock}{\relax}
\providecommand{\bibinfo}[2]{#2}
\providecommand{\BIBentrySTDinterwordspacing}{\spaceskip=0pt\relax}
\providecommand{\BIBentryALTinterwordstretchfactor}{4}
\providecommand{\BIBentryALTinterwordspacing}{\spaceskip=\fontdimen2\font plus
\BIBentryALTinterwordstretchfactor\fontdimen3\font minus
  \fontdimen4\font\relax}
\providecommand{\BIBforeignlanguage}[2]{{%
\expandafter\ifx\csname l@#1\endcsname\relax
\typeout{** WARNING: IEEEtran.bst: No hyphenation pattern has been}%
\typeout{** loaded for the language `#1'. Using the pattern for}%
\typeout{** the default language instead.}%
\else
\language=\csname l@#1\endcsname
\fi
#2}}
\providecommand{\BIBdecl}{\relax}
\BIBdecl

\bibitem{grotz2010}
J.~Grotz, B.~Ottersten, and J.~Krause, ``Signal detection and synchronization
  for interference overloaded satellite broadcast reception,'' \emph{{IEEE}
  Trans. Wireless Commun.}, vol.~9, no.~10, pp. 3052 --3063, October 2010.

\bibitem{hicks2001}
J.~Hicks, S.~Bayram, W.~Tranter, R.~Boyle, and J.~Reed, ``Overloaded array
  processing with spatially reduced search joint detection,'' \emph{{IEEE} J.
  Sel. Areas Commun.}, vol.~19, no.~8, pp. 1584 --1593, aug 2001.

\bibitem{kapur2003}
A.~Kapur and M.~Varanasi, ``Multiuser detection for overloaded {CDMA}
  systems,'' \emph{{IEEE} Trans. Inf. Theory}, vol.~49, no.~7, pp. 1728 --
  1742, july 2003.

\bibitem{Colman2008}
G.~Colman and T.~Willink, ``Overloaded array processing using genetic
  algorithms with soft-biased initialization,'' \emph{{IEEE} Trans. Veh.
  Technol.}, vol.~57, no.~4, pp. 2123 --2131, July 2008.

\bibitem{krause2011}
M.~Krause, D.~Taylor, and P.~Martin, ``List-based group-wise symbol detection
  for multiple signal communications,'' \emph{{IEEE} Trans. Wireless Commun.},
  vol.~10, no.~5, pp. 1636 --1644, may 2011.

\bibitem{beidas2002}
B.~Beidas, H.~El~Gamal, and S.~Kay, ``Iterative interference cancellation for
  high spectral efficiency satellite communications,'' \emph{{IEEE} Trans.
  Commun.}, vol.~50, no.~1, pp. 31 --36, jan 2002.

\bibitem{Schwarz2007}
K.~Schwarzenbarth, J.~Grotz, and B.~Ottersten, ``{MMSE} based interference
  processing for satellite broadcast reception,'' in \emph{Proc. Vehicular
  Tech. Conf.}, Apr. 2007, pp. 1345 --1349.

\bibitem{honig2008}
M.~L. Honig, ``Overview of multiuser detection,'' in \emph{Advances in
  Multiuser Detection}.\hskip 1em plus 0.5em minus 0.4em\relax M. L. Honig, ed.
  Hoboken, NJ, USA: John Wiley \& Sons, Inc., 2008, pp. 1--45.

\bibitem{dvbb-s2}
ETSI, ``{EN} 302 307: {D}igital {V}ideo {B}roadcasting {(DVB)}; second
  generation framing structure, channel coding and modulation systems for
  broadcast, interactive services, news gathering and other broadband satellite
  applications,'' \emph{ETSI}, 2005.

\bibitem{eigen}
T.~De~Bie, N.~Cristianini, and R.~Rosipal, ``Eigenproblems in pattern
  recognition,'' in \emph{Handbook of Geometric Computing: Applications in
  Pattern Recognition, Computer Vision, Neural computing, and Robotics}.\hskip
  1em plus 0.5em minus 0.4em\relax E. B. Corrochano, ed. Berlin Heidelberg,
  Germany: Springer, 2005, pp. 128--132.

\bibitem{Verdu:1998:MD:521411}
S.~Verdu, \emph{Multiuser Detection}, 1st~ed.\hskip 1em plus 0.5em minus
  0.4em\relax New York, NY, USA: Cambridge University Press, 1998.

\end{thebibliography}
\end{document}